\documentclass[journal]{IEEEtran}
\IEEEoverridecommandlockouts

\usepackage{algorithm,algorithmic,amsmath,amssymb,amsthm,bbm,cite,color,comment,graphicx,microtype,setspace,subfigure,url,xcolor}
\usepackage[nolist]{acronym}
\usepackage[USenglish]{babel}
\usepackage[T1]{fontenc}
\usepackage[utf8]{inputenc}
\usepackage{hyperref}
\usepackage[utf8]{inputenc}
\usepackage{newunicodechar}
\newunicodechar{Κ}{K}
\usepackage[figurename=Fig., labelsep=period, font=small]{caption}

\usepackage{tikz,pgfplots}
\usetikzlibrary{arrows}
\pgfplotsset{compat=newest}
\pgfplotsset{
/pgfplots/ybar legend/.style={
/pgfplots/legend image code/.code={%
\draw[##1,/tikz/.cd,bar width=0.1cm,yshift=-0.2em,bar shift=0.5*\pgfplotbarwidth]
plot coordinates {(0.5*\pgfplotbarwidth,0.6em) (2.5*\pgfplotbarwidth,0.4em) (4.5*\pgfplotbarwidth,0.2em)};},}}

\usepackage{titlesec}
\let\subparagraph\relax
\titlespacing{\section}{0pt}{6pt plus 2pt minus 1pt}{4pt plus 1pt minus 1pt} 
\titlespacing{\subsection}{0pt}{4pt plus 2pt minus 1pt}{2pt plus 1pt minus 1pt} 
\graphicspath{{figures/}}
\usepackage{textcomp}
\def\BibTeX{{\rm B\kern-.05em{\sc i\kern-.025em b}\kern-.08em
    T\kern-.1667em\lower.7ex\hbox{E}\kern-.125emX}}

\usepackage{changepage}
\usepackage{calc}

\usepackage{multibib}
\newcites{resp}{References for Response Letter}

\usepackage{todonotes}
\renewcommand{\b}{\mathbf{b}}

\newcommand{\h}{\mathbf{h}}

\newcommand{\n}{\mathbf{n}}

\newcommand{\p}{\mathbf{p}}
\newcommand{\q}{\mathbf{q}}

\newcommand{\w}{\mathbf{w}}

\newcommand{\y}{\mathbf{y}}

\newcommand{\0}{\mathbf{0}}

\newcommand{\B}{\mathbf{B}}

\newcommand{\I}{\mathbf{I}}

\newcommand{\N}{\mathbf{N}}

\newcommand{\R}{\mathbf{R}}

\newcommand{\U}{\mathbf{U}}

\newcommand{\W}{\mathbf{W}}

\newcommand{\Y}{\mathbf{Y}}

\newcommand{\setC}{\mathcal{C}}

\newcommand{\setN}{\mathcal{N}}
\newcommand{\setO}{\mathcal{O}}

\newcommand{\setCN}{\mathcal{CN}}

\newcommand{\Compl}{\mbox{$\mathbb{C}$}}

\newcommand{\herm}{\mathrm{H}}

\newcommand{\tran}{\mathrm{T}}

\definecolor{oulu_blue}{HTML}{23408F}
\definecolor{oulu_green}{HTML}{39B54A}

\definecolor{red}{rgb}{1,0,0}
\definecolor{red_magenta}{rgb}{1,0,0.5}
\definecolor{magenta}{rgb}{1,0,1}
\definecolor{blue_magenta}{rgb}{0.5,0,1}
\definecolor{blue}{rgb}{0,0,1}
\definecolor{blue_cyan}{rgb}{0,0.5,1}
\definecolor{cyan}{rgb}{0,0.75,0.75} 
\definecolor{green_cyan}{rgb}{0,1,0.5}
\definecolor{green}{rgb}{0,1,0}
\definecolor{green_yellow}{rgb}{0.375,0.75,0} 
\definecolor{yellow}{rgb}{1,1,0}
\definecolor{red_yellow}{rgb}{1,0.5,0}

\begin{acronym}
\acro{2D-MUSIC}{two-dimensional multiple signal classification}
\acro{AWGN}{additive white Gaussian noise}
\acro{BS}{base station}
\acro{CSI}{channel state information}
\acro{DoF}{degree of freedom}
\acro{i.i.d.}{independent and identically distributed}
\acro{LoS}{line-of-sight}
\acro{LS}{least-squares}
\acro{MF}{matched filtering}
\acro{MIMO}{multiple-input multiple-output}
\acro{MUSIC}{multiple signal classification}
\acro{mmWave}{millimeter-wave}
\acro{NLoS}{non-line-of-sight}
\acro{PDF}{probability density function}
\acro{SE}{spectral efficiency}
\acro{SINR}{signal-to-interference-plus-noise ratio}
\acro{SNR}{signal-to-noise ratio}
\acro{sub-THz}{sub-terahertz}
\acro{sum-SE}{sum spectral efficiency}
\acro{ULA}{uniform linear array}
\acro{UL}{uplink}
\acro{ZF}{zero forcing}
\end{acronym}

\title{Localization-Based Beam Focusing in Near-Field Communications}
\author{Nima Mozaffarikhosravi, Prathapasinghe Dharmawansa, and Italo Atzeni
\thanks{The authors are with the Centre for Wireless Communications, University of Oulu, Finland (e-mail: \{nima.mozaffari, prathapasinghe.kaluwadevage, italo.atzeni\}@oulu.fi). This work was supported by the Research Council of Finland (336449 Profi6, 348396 HIGH-6G, and 369116 6G~Flagship).} \vspace{-0.5mm}
}

\begin{document}

\maketitle

\begin{abstract}
Shifting 6G-and-beyond wireless systems to higher frequency bands and the utilization of massive multiple-input multiple-output arrays will extend the near-field region, affecting beamforming and user localization schemes. In this paper, we propose a localization-based beam-focusing design, in which the receive combiners are directly constructed from the steering vectors corresponding to the estimated user locations. To support this approach, we analyze the 2D-MUSIC algorithm by examining its spectrum in simplified, tractable setups with minimal numbers of antennas and users. Lastly, we compare the proposed localization-based beam focusing, with locations estimated via 2D-MUSIC, with pilot-based zero forcing in terms of uplink sum spectral efficiency. Our results show significant gains under dominant line-of-sight propagation, short coherence blocks, and high noise power typical of high-frequency systems.
\end{abstract} \vspace{-1.5mm}

\vspace{-0.5mm}

\section{Introduction}\label{Introduction}

Emerging wireless applications such as holographic communications and high-precision sensing are expected to introduce new bandwidth demands in 6G-and-beyond systems. A promising way to accommodate this growth is to shift operations toward higher frequency bands, including the \ac{mmWave} and \ac{sub-THz} ranges \cite{Italo2025_THz}. As the carrier frequency increases, massive \ac{MIMO} arrays are needed to overcome the challenging propagation; at the same time, antennas become physically smaller, thus facilitating the deployment of compact massive \ac{MIMO} arrays \cite{Emil2017_mMIMO}. When operating with large arrays at smaller wavelengths, the radiative near-field region extends significantly, affecting both beamforming (or beam focusing) and user localization schemes \cite{bodet2024_sub}. The near-field spherical wavefronts can be leveraged to focus the signals on precise spatial locations, enabling user localization by estimating both distance and angle with a single \ac{BS}. In addition, due to increased pathloss and scattering absorption at \ac{mmWave} and \ac{sub-THz} frequencies, \ac{NLoS} paths become less impactful and the channel is typically dominated by \ac{LoS} components \cite{Italo2025_THz}. Thus, it is more practical to exploit user localization methods for beam-focusing design, as their comparatively lower signaling overhead can help enhance the system's \ac{SE} \cite{Emil2025_Loc&Sense}.

In this regard, \cite{Ram2025_loc} compared far-field and near-field user localization and reviewed the most commonly used algorithms in the literature. Moreover, \cite{Lei2025_Loc&Est} examined the challenges of user localization and channel estimation in the near field, also surveying relevant algorithms. The study in \cite{Emil2025_MUSICAnalysis} analyzed the \ac{2D-MUSIC} algorithm for parametric near-field channel estimation in multi-user \ac{MIMO} systems, demonstrating that \ac{2D-MUSIC} achieves near-optimal performance compared to non-parametric and sequential alternatives. Furthermore, \cite{Ram2024_ModifiedMUSIC} proposed a modification of the \ac{2D-MUSIC} algorithm that reduces the complexity of the search procedure, therefore facilitating its application in localization-based near-field communications. Recently, \cite{gavras2024_nearfieldLoc} presented a near-field localization approach for a hybrid analog-digital receiver equipped with a dynamic metasurface antenna and 1-bit analog-to-digital converters. While prior works have investigated near-field localization and channel estimation, the explicit use of localization for beam-focusing design, particularly leveraging the \ac{LoS}-dominated propagation at high frequencies, remains unaddressed.

In this paper, we consider a near-field channel model with \ac{LoS} and \ac{NLoS} propagation and provide the following contributions. We propose a localization-based \mbox{beam-focusing} design, in which the receive combiners are directly constructed from the steering vectors corresponding to the estimated user locations. To support this approach, we analyze the \ac{2D-MUSIC} algorithm in simplified setups with minimal numbers of antennas and users, enabling tractable analysis of the distance estimation error and interference effects. Lastly, we compare the proposed localization-based beam focusing, with locations estimated via \ac{2D-MUSIC}, with pilot-based \ac{ZF}, evaluating the \ac{UL} \ac{sum-SE} across various system and design parameters. In this comparison, we also incorporate a frequency-dependent reflection coefficient that accounts for the well-established \ac{LoS}-dominated propagation at high carrier frequencies, modeled based on real measurement data. Our approach yields significant gains under \ac{LoS}-dominated propagation, short coherence blocks, and high noise power typical of high-frequency systems, while remaining effective in the presence of residual~\ac{NLoS}~components.

\section{System and Channel Model}

In this section, we introduce the considered system model and near-field channel model with \ac{LoS} and \ac{NLoS} propagation and frequency-dependent reflection.

\subsection{System Model} \label{System Model}

Consider a \ac{MIMO} system where $K$ single-antenna users are served by a \ac{BS} with $N$ antennas in the \ac{UL}. The received data signal at the \ac{BS} is given by
\begin{align}\label{recieved_signal}
\y = \sum_{k=1}^{K} \sqrt{\rho_{k}}\h_{k} {s}_k + \n \in \mathbb{C}^{N \times 1},
\end{align}
where $\rho_{k}$ represents the transmit power of user~$k$, $\h_k \in \mathbb{C}^{N \times 1}$ is the random channel between user~$k$ and the \ac{BS}, ${s}_k \in \Compl$ denotes the data symbol of user~$k$ satisfying $\mathbb{E}[|s_{k}|^2] = {1}$ and $\mathbb{E}[{s}_k{s}_i^\mathrm{*}] ={0}$, $\forall i \neq k$, and $\n \sim \setCN (\0,\sigma^2\I_{N})$ is a vector of \ac{AWGN}. Considering a block-fading model, the channels are constant over a coherence block of length $T$. Moreover, we assume that the users are located in the near field of the \ac{BS}, as detailed in Section~\ref{Near-Field Channel Model}.

Channel estimation in time-division duplexing is typically carried out via \ac{UL} pilots, transmitted simultaneously by all the users. Let $\p_{k} \in \mathbb{C}^{\tau_\textrm{Pil} \times 1}$ denote the pilot of user~$k$, where $\tau_\textrm{Pil}$ is the pilot length. Assuming $\tau_\textrm{Pil}\geq K$, we consider orthogonal pilots such that $\p_{k}^\herm \p_{i}=0$, $\forall i \neq k$ and we assume that each entry of $\p_k$ has unit magnitude to ensure a constant power level. The received pilot signal at the \ac{BS} is given by 
\begin{align}\label{recieved_signal_PILOT}
\Y_\textrm{Pil} = \sum_{k=1}^{K} \sqrt{\rho_{k}}\h_{k} \p_k^\tran + \N_\textrm{Pil} \in \mathbb{C}^{N \times \tau_\textrm{Pil}},
\end{align}
where $\N_\textrm{Pil} \in \mathbb{C}^{N \times \tau_\textrm{Pil}}$ is a matrix of \ac{AWGN} with \ac{i.i.d.} $\setC \setN (0,\sigma^2)$ entries. The channel estimate is then obtained from \eqref{recieved_signal_PILOT} using, for example, the \ac{LS} estimator $\hat{\h}_k = \frac{1}{\sqrt{\rho_{k}}\tau_\textrm{Pil}}\Y_\textrm{Pil}\p_{k}^* \in \mathbb{C}^{N \times 1}$, which does not require any statistical \ac{CSI} and is considered as part of our baseline in Section~\ref{sec:SE_evaluation}.

Let $\w_k \in \mathbb{C}^{N \times 1}$ denote the receive combiner corresponding to user~$k$. The resulting \ac{SINR} of user~$k$ is given by
\begin{align}\label{SINR}
 \text{SINR}_{k} = \frac{\rho_{k} |\mathbf{w}_{k}^\herm\mathbf{h}_{k}|^2}{
 \sum_{i \neq k}
 \rho_{i} | \w_{k}^\herm \h_{i} |^2 + \sigma^2 \| \w_{k}\|^2}.
\end{align}
Finally, the \ac{UL} \ac{sum-SE} of the system can be expressed as $\text{sum-SE} = \big(1-\frac{\tau_\textrm{Pil}}{T}\big) \mathbb{E} \big[\sum_{k=1}^{K}\log_2 (1 + \text{SINR}_{k})\big]$ [bit/s/Hz], where the expected value is over the channel realizations. While the receive combiners are typically constructed based on the channel estimates, we propose an alternative localization-based receive combiner design in Section~\ref{beamfocusing}, which turns out to be remarkably effective under \ac{LoS}-dominated propagation.

\subsection{Near-Field Channel Model} \label{Near-Field Channel Model}

Assume that the \ac{BS} is equipped with a \ac{ULA} with antenna spacing $d$ and total length $D = (N-~1) d$. Denote the carrier frequency as $f_\textrm{c}$ and the corresponding wavelength as $\lambda_\textrm{c} =\frac{c}{f_\textrm{c}}$, where $c$ is the speed of light in vacuum. The Fraunhofer distance, which is commonly used to separate the near- and far-field regions, is defined as $R_\textrm{F} = \frac{2D^2}{\lambda_\textrm{c}}$. We assume that the carrier frequency, the \ac{ULA} length, and the distance of the users and scatterers from the \ac{BS} are such that both the \ac{LoS} and \ac{NLoS} paths are subject to near-field propagation. For a source defined by the distance $r$ and angle $\theta$, the near-field steering vector is given by \cite{Dai2022_TCOMChannel}
\begin{align}\label{steering vector}
\b(f_\textrm{c}, r, \theta) = \begin{bmatrix} e^{-j \frac{2 \pi }{\lambda_\textrm{c}} (\bar{r}_0 (r, \theta) - r)} \vspace{-2mm} \\ \vdots \\ e^{-j \frac{2 \pi}{\lambda_\textrm{c}} (\bar{r}_{N-1} (r, \theta) - r)} \end{bmatrix} = \begin{bmatrix} b_{1} \vspace{-1mm} \\ \vdots \\ b_{N} \end{bmatrix} \in \Compl^{N \times 1},
\end{align}
where $\bar{r}_{n} (r, \theta) = \sqrt{r^2 + \delta_n^2 d^2 - 2 r \delta_n d \sin \theta}$ denotes the distance from the $n$th antenna, with $\delta_n =\frac{2n-N+1}{2}$, $\forall n = 0, \ldots, N-1$. For notational simplicity, we omit the dependence of $\bar{r}_{n}$ on $r$ and $\theta$ in the following.

We consider a general near-field channel model with \ac{LoS} and \ac{NLoS} propagation, where the latter follows the clustered multipath model with $L$ \ac{NLoS} paths \cite{Dai2022_TCOMChannel}, \cite[Ch.~5.6.1]{Emil2024_MultiAntenna}. Each \ac{NLoS} path originates from a cluster comprising a large number of scatterers, which contributes to the signal’s random phase rotation and power fluctuation. Without loss of generality, we assume that the $L$ clusters are common to all the users. Let $\theta_{k} \in \big[-\frac{\pi}{2}, \frac{\pi}{2}\big]$ and $\theta_{\ell}^{\textrm{NLoS}} \in \big[-\frac{\pi}{2}, \frac{\pi}{2}\big]$ denote the angles of user~$k$ and the $\ell$th cluster, respectively, relative to the \ac{ULA}’s broadside direction. Furthermore, let $r_{k}$ and $r_{\ell}^{\textrm{NLoS}}$ denote the distances from user~$k$ and the $\ell$th cluster, respectively, to the center of the \ac{ULA}. Considering a common pathloss factor for all the antennas as in \cite{Dai2022_TCOMChannel}, the channel of user~$k$ is modeled as
\begin{align} \label{channel}
    \mathbf{h}_k = \underbrace{\frac{\lambda_\textrm{c}}{4\pi r_k} e^{-j\frac{2\pi}{\lambda_\textrm{c}}r_k}\mathbf{b}(f_\textrm{c}, r_k, \theta_k)}_{\textrm{LoS}} + \underbrace{\sum_{\ell=1}^{L} g_{k, \ell} \mathbf{b}(f_\textrm{c},r_{\ell}^{\textrm{NLoS}}, \theta_{\ell}^{\textrm{NLoS}}) }_{\textrm{NLoS}},
\end{align}
where $g_{k,\ell}$ is the small-scale fading coefficient of user~$k$'s $\ell$th \ac{NLoS} path. Note that, for the \ac{NLoS} paths, we have ignored the first hop from the users to the clusters to avoid making assumptions about the clusters' geometry and relative distances of the users: this results in a slight overestimation of the power of the \ac{NLoS} paths, which in turn underestimates the gains of the proposed localization-based beam focusing. Hence, we have $g_{k,\ell} \sim \setCN (0,\gamma_{\ell}^2)$, $\forall k=1,\ldots,K$, where $\gamma_{\ell}^2$ models the large-scale fading of the $\ell$th \ac{NLoS} path. In this paper, we introduce an additional frequency-dependent reflection coefficient, which is typically not included in existing works. Specifically, we~define
\begin{align}
\gamma_{\ell}^2 = \bigg(\frac{\lambda_\textrm{c}}{4\pi r_{\ell}^{\textrm{NLoS}}}\bigg)^2 \alpha_{\ell}(f_\textrm{c}),
\end{align}
where $\alpha_{\ell} (f_{\textrm{c}}) \in (0,1]$ models the reflection at the $\ell$th scatterer. This term captures the increased scattering absorption at higher frequencies and is thus a decreasing function of $f_{\textnormal{c}}$. Notably, setting $\alpha_{\ell} (f_{\textnormal{c}}) = 1$ recovers existing channel models (e.g., \cite{Dai2022_TCOMChannel}) that assume perfect reflection. In Section~\ref{sec:SE_evaluation}, we adopt a function for $\alpha_{\ell} (f_{\textnormal{c}})$ based on real measurement data from \cite{Con2025_Absorption}.

\section{Localization-Based Beam Focusing} \label{beamfocusing}

Motivated by the well-established fact that the \ac{NLoS} paths tend to become less impactful at higher frequencies due to the increased pathloss and scattering absorption, we propose a localization-based beam-focusing strategy that leverages \ac{LoS}-dominated propagation and maintains solid performance in the presence of residual \ac{NLoS} components. In the proposed approach, the receive combiners are obtained by replacing the estimated channels $\{ \hat{\h}_{k} \}$ with $\big\{ \b (f_\textrm{c}, \hat{r}_{k}, \hat{\theta}_{k}) \big\}$ in conventional beamforming structures, where $\hat{r}_{k}$ and $\hat{\theta}_{k}$ are the estimated distance and angle, respectively, of user~$k$. In this paper, we consider a \ac{ZF} structure with receive combining matrix given~by
\begin{align} \label{eq:BF}
\W = [\w_{1}, \ldots, \w_{K}] = \hat{\B} (\hat{\B}^\herm \hat{\B})^{-1} \in \Compl^{N \times K},
\end{align}
with $\hat{\B} = \big[\b (f_\textrm{c}, \hat{r}_{1}, \hat{\theta}_{1}), \ldots, \b (f_\textrm{c}, \hat{r}_{K}, \hat{\theta}_{K})\big] \in \mathbb{C}^{N \times K}$. In the rest of this section, we describe near-field user localization based on the \ac{2D-MUSIC} algorithm, with emphasis on distance estimation. The \ac{sum-SE} performance of the proposed localization-based beam focusing is evaluated in Section~\ref{sec:SE_evaluation} as a function of various system and design parameters.

The \ac{2D-MUSIC} algorithm leverages the orthogonality between the signal and noise subspaces to localize the users via a 2D grid search for distance and angle \cite{Ram2024_ModifiedMUSIC}. First, \ac{UL} pilots similar to those in Section~\ref{System Model} are simultaneously transmitted by the users. Let $\q_{k} \in \mathbb{C}^{\tau_\textrm{Loc} \times 1}$ denote the localization pilot for user~$k$, where $\tau_{\textrm{Loc}}$ is the pilot length. The received localization pilot signal at the \ac{BS} is given by (cf. \eqref{recieved_signal_PILOT})
\begin{align}\label{recieved_signal_loc}
\Y_\textrm{Loc} = \sum_{k=1}^{K} \sqrt{\rho_{k}}\h_{k} \q_k^\tran + \N_\textrm{Loc} \in \Compl^{N \times \tau_{\textrm{Loc}}},
\end{align}
where $\N_\textrm{Loc} \in \mathbb{C}^{N \times \tau_\textrm{Loc}}$ is a matrix of \ac{AWGN} with \ac{i.i.d.} $\setC \setN (0,\sigma^2)$ entries.
Based on \eqref{recieved_signal_loc}, the sample covariance matrix of the received signal is constructed as
\begin{align} \label{cov_matrix_loc}
\R_{\textrm{Loc}} = \frac{1}{\tau_{\textrm{Loc}}}\Y_{\textrm{Loc}} \Y_{\textrm{Loc}}^{\herm} \in \Compl^{N \times N}.
\end{align}
Assuming the eigenvalues of $\R_{\textrm{Loc}}$ are sorted in descending order, the signal and noise subspaces are approximately spanned by the first $K$ and last $N-K$ eigenvectors of $\R_{\textrm{Loc}}$, respectively. In this regard, $\tau_{\textrm{Loc}}$ governs how accurately $\R_{\textrm{Loc}}$ captures these subspaces, and a small value of $\tau_{\textrm{Loc}}$ (much smaller than $\tau_{\textrm{Pil}}$) typically suffices~\cite{Lei2025_Loc&Est}: this benefits the effective \ac{sum-SE}, as shown in Section~\ref{sec:SE_evaluation}. Let $\U_\textrm{n} \in \mathbb{C}^{N \times (N-K)}$ denote the noise eigenvector matrix comprising the last $N-K$ eigenvectors of $\R_{\textrm{Loc}}$. Then, the \ac{2D-MUSIC} algorithm searches for the steering vectors most orthogonal to $\U_\textrm{n}$ by evaluating the spectrum
\begin{align}\label{spectrum}
S(f_\textrm{c}, r, \theta) = \frac{1}{\b^\herm(f_\textrm{c}, r, \theta) \U_\textrm{n} \U_\textrm{n}^\herm \b(f_\textrm{c}, r, \theta)}
\end{align}
over a grid of $(r, \theta)$ values, and the $K$ largest peaks yield the estimated locations of the $K$ users \cite{Ram2024_ModifiedMUSIC}. Note that the search resolution in $(r,\theta)$ affects both the localization accuracy and complexity, as discussed in Section~\ref{sec:SE_evaluation}.

To investigate the distinctive features of the \ac{2D-MUSIC} algorithm in terms of distance estimation error and impact of interference, we analyze its spectrum in simplified setups with minimal numbers of antennas and users. These configurations allow for tractable analysis and offer deeper insights into the algorithm's behavior. Nevertheless, the resulting conclusions extend to more general scenarios, as confirmed by our numerical results. To accurately characterize the algorithm's performance in terms of distance estimation, we assume perfectly known angles in the remainder of this section.

\subsection{Distance Estimation Error} \label{sec:MUSIC_1}

To analyze the distance estimation error, we begin by considering a simplified setup with $K = 1$ and $N = 2$, which represents the simplest scenario that enables distance estimation with the \ac{2D-MUSIC} algorithm. For this setup with no interference, we can derive a closed-form expression for the estimated distance by maximizing \eqref{spectrum} (i.e., minimizing its denominator). We begin by writing the denominator of \eqref{spectrum} as
\begin{align}
    \frac{1}{S(f_\textrm{c}, r,\theta)} & = |u_1 b_1^* + u_2 b_2^*|^2 \notag = |u_1|^2 + |u_2|^2 +  2|u_1||u_2| \notag \\
    &  \phantom{=} \ \times \cos\bigg( \frac{2\pi}{\lambda_\textrm{c}}(\bar{r}_0 - \bar{r}_1) + \phi_1 - \phi_2 \bigg), \label{deno2antenna}
\end{align}
with $\U_\textrm{n} = [u_1 \ u_2]^\tran \in \Compl^{2 \times 1}$, $\bar{r}_0 = \sqrt{r^2 + \frac{d^2}{4} + rd \sin\theta}$, and $\bar{r}_1 = \sqrt{r^2 + \frac{d^2}{4} - rd \sin\theta}$, and where $\phi_k = \angle u_k$ denotes the phase of $u_k$, $\forall k \in \{1, 2\}$. The cosine term is minimized at $-1$ when $|u_1|, |u_2| > 0$ and $\theta$ is known, resulting in

$ $

\vspace{-8mm}

\begin{align}\label{cosine_min}
\frac{2\pi}{\lambda_\textrm{c}}(\bar{r}_0 - \bar{r}_1) 
+ \phi_1 - \phi_2 
&= (2M+1)\pi, \quad M \in \mathbb{Z}.
\end{align}
Plugging in the expressions of $\bar{r}_0$ and $\bar{r}_1$ and squaring \eqref{cosine_min}~gives
\begin{align}
2r^2 + \kappa(\phi_1,\phi_2) &= 2\sqrt{\,r^4 + \tfrac{r^2 d^2}{2} + \tfrac{d^4}{16} - r^2 d^2 \sin^2\theta},
\label{eq:pre_square}
\end{align}
with $\kappa(\phi_1, \phi_2)=\tfrac{d^2}{2} - \big(\tfrac{\lambda_\textrm{c}}{2 \pi}\big((2M+1)\pi - \phi_1 + \phi_2\big)\big)^2$, $M \in \mathbb{Z}$. Finally, squaring \eqref{eq:pre_square} again and collecting terms in $r^2$ leads to
\begin{align}\label{closed_form}
\hat{r}_1^{\,2} &= \frac{\tfrac{d^4}{4} - \kappa^2(\phi_1,\phi_2)} {4\kappa(\phi_1,\phi_2) - 2d^2 + 4d^2\sin^2\theta}.
\end{align}
However, \eqref{closed_form} admits multiple values of $M$ as \eqref{deno2antenna} oscillates due to the presence of the cosine term, leading to multiple possible minimizing distances. The choice of $M$ associated with the accurate estimated distance can only be determined numerically around the true distance, assuming that a rough estimate of the true location is available. To circumvent this limitation, we propose a bound on $M$ that is remarkably tight for all cases of our interest. Using the factorization $x^2-y^2=(x-y)(x+y)$, $x,y\in\mathbb{C}$, we rewrite \eqref{closed_form} as
\begin{align}
\hat{r}_{1}^2 
= \frac{\Psi^2 (d^2 - \Psi^2)}{4(d^2 \sin^2 \theta - \Psi^2)} = \frac{\Psi^2 (d+\Psi)}{4(d\sin\theta + \Psi)} \frac{\tfrac{\Psi}{d}-1}{\tfrac{\Psi}{d}-\sin\theta},
\end{align}
with $\Psi = \tfrac{\lambda_\textrm{c}}{2 \pi}\big((2M+1)\pi - \phi_1 + \phi_2\big)$. Now, keeping in mind that $\hat{r}_1>0$, we have that $\frac{\Psi}{d} \geq 1$, with $\phi_2-\phi_1\in(0,2\pi]$. Since the exact distribution of $\phi_2-\phi_1$ is intimately related to the eigenvector distribution of correlated non-central Wishart matrices (which remains unknown even for the $2\times 2$ case), we may take $\pi$ as a rough estimate of its mean to yield
\begin{align} \label{bound_M}
    M \geq \frac{d}{\lambda_\textrm{c}} - 1.
\end{align}

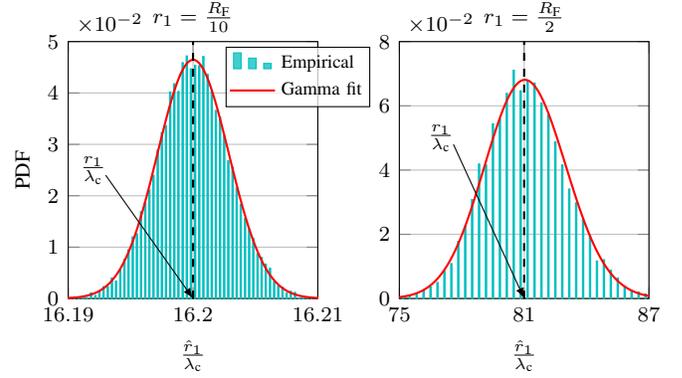
\begin{figure}[t!]
    \centering
        \begin{tikzpicture}
            \begin{axis}[
                width=4.9cm, height=5cm,
                xmin=16.19, xmax=16.21,
                ymin=0, ymax=0.05,
                xlabel={$\frac{\hat{r}_{1}}{\lambda_\textrm{c}}$},
                ylabel={PDF},
                label style={font=\footnotesize},
                xtick={16.19,16.2,16.21},
                ytick={0,0.01,...,0.05},
                tick scale binop=\times,
                ticklabel style={font=\footnotesize},
                legend style={at={(1.21,0.98)}, anchor=north east, font=\scriptsize, inner sep=1pt, fill opacity=0.75, draw opacity=1, text opacity=1},
                legend cell align=left,
                title={$r_{1} = \frac{R_\textrm{F}}{10}$},
                title style={font=\footnotesize, yshift=-2mm},
                grid=major,
                ]
                
                \addplot[ybar, bar width=0.5pt, draw=cyan, fill=cyan, ybar legend]
                table[x index=0, y index=1, col sep=space]
                {figures/gamma_hist_df_over_10.tex};

                \addplot[red, thick]
                table[x index=0, y index=1, col sep=space]
                {figures/gamma_dist_df_over_10.tex};
                 
                \addplot[black, dashed, thick] coordinates {(16.2, 0) (16.2, 0.05)};

                \begin{scope}[>=latex]
                \draw[->] (axis cs:16.193,0.024) -- (axis cs:16.2,0);
                \end{scope}

                \node[black, font=\footnotesize] at (axis cs:16.192,0.025) {$\frac{r_{1}}{\lambda_\textrm{c}}$};

                \legend{Empirical, Gamma fit}
            \end{axis}
        \end{tikzpicture} \hspace{-2.5mm}
        \begin{tikzpicture}
            \begin{axis}[
                width=4.9cm, height=5cm,
                xmin=75, xmax=87,
                ymin=0, ymax=0.08,
                xlabel={$\frac{\hat{r}_{1}}{\lambda_\textrm{c}}$},
                label style={font=\footnotesize},
                xtick={75,81,87},
                ytick={0,0.02,...,0.08},
                tick scale binop=\times,
                ticklabel style={font=\footnotesize},
                title={$r_{1} = \frac{R_\textrm{F}}{2}$},
                title style={font=\footnotesize, yshift=-2mm},
                grid=major,
                ]
                
                \addplot[ybar, bar width=0.5pt, draw=cyan, fill=cyan, ybar legend]
                table[x index=0, y index=1, col sep=space]
                {figures/gamma_hist_df_over_2.tex};

                \addplot[red, thick]
                table[x index=0, y index=1, col sep=space]
                {figures/gamma_dist_df_over_2.tex};
                
                \addplot[black, dashed, thick] coordinates {(81, 0) (81, 0.08)};
        
                \begin{scope}[>=latex]
                \draw[->] (axis cs:77.6,0.048) -- (axis cs:81,0);
                \end{scope}

                \node[black, font=\footnotesize] at (axis cs:77,0.05) {$\frac{r_{1}}{\lambda_\textrm{c}}$};

            \end{axis}
        \end{tikzpicture}
        
    \caption{Empirical PDF of the estimated distance based on \eqref{closed_form}, with $K=1$, $N=2$, $f_\textrm{c}= 100$~GHz, $\textrm{SNR} = 20$~dB, $d = 9 \lambda_\textrm{c}$, and $\theta_1 = 0$.} \vspace{-1mm}
    \label{hist}
\end{figure}

Fig.~\ref{hist} plots the empirical \ac{PDF} of the estimated distance based on \eqref{closed_form}, with $f_\textrm{c} = 100$~GHz, $\textrm{SNR} = 20$~dB, $d = 9 \lambda_\textrm{c}$, and $r_{1} \in \big\{ \frac{R_\textrm{F}}{10}, \frac{R_\textrm{F}}{2} \big\}$. Here, \eqref{bound_M} gives $M \geq 8$ against an exact value of $M=8$, further confirming the tightness of the proposed bound. Notably, $\hat{r}_{1}$ can be approximately modeled as a Gamma random variable (see, e.g., \cite{Al-Ahm2010_GammaDist} and references therein), i.e., $\hat{r}_{1} \sim \mathcal{G}(\mu, \nu)$, with
\begin{align} \label{eq:r_gamma_definition}
    \mathbb{E}[\hat{r}_{1}] \simeq r_{1}, \quad 
    \mathrm{Var}[\hat{r}_{1}] = \zeta^2 \simeq \mu \nu^2,
\end{align}
where $\mu$ and $\nu$ are the shape and scale parameters, respectively. Although it is intractable to analytically derive the statistical properties of $\hat{r}_{1}$ (i.e., $\mu$ and $\nu$) under the \ac{2D-MUSIC} algorithm, our simulation results show that the variance of the estimated distance grows as $\setO (r_1^4)$, consistent with the Cramér-Rao bound analysis in \cite{Korso2010_CRB}. This scaling can be interpreted geometrically using the Fresnel approximation: for $\theta_{1} = 0$, the distance from an arbitrary antenna~$n$ satisfies $\bar{r}_{n}(r_1, 0) \simeq r_1 + \frac{\delta_n^2 d^2}{2r_1}$. Taking the derivative of the second term with respect to $r_1$ captures how the curvature varies with the distance, showing a dependence proportional to $\frac{1}{r_1^2}$. Since the estimation variance scales with the inverse square of this term, it grows as $\setO (r_1^4)$. This trend is plotted in Fig.~\ref{Error_STD} considering $K = 1$, $N \in \{ 64, 128 \}$, $f_\textrm{c} = 100$~GHz, $\textrm{SNR}=20$~dB, and $d = \frac{\lambda_\textrm{c}}{2}$. Hence, the variance can be accurately approximated as
\begin{align} \label{eq:zeta}
    \zeta^2(r_{1}) \simeq \eta r_{1}^4,
\end{align}
where $\eta>0$ can be obtained numerically by curve fitting on simulated data. This scaling property turns out to be independent of both the number of antennas and their spacing. Thus, the variance of the estimated distance for any point in the near field can be approximated based on the numerically evaluated variance at a given reference distance.

\begin{figure}[t!]
    \centering
    \begin{tikzpicture}
    \begin{axis}[
        width=4.25cm, height=5cm,
        ymin=0, ymax=0.1,
        xmin=0.59535, xmax=2.97675,
        xlabel={${r}_{1} $},
        ylabel={$\zeta^2(r_{1})$},
        label style={font=\footnotesize},
        xtick={2.97675, 1.9845, 1.488375, 0.992252, 0.59535},
        xticklabels={$\frac{R_\textrm{F}}{2}$, $\frac{R_\textrm{F}}{3}$, $\frac{R_\textrm{F}}{4}$, $\frac{R_\textrm{F}}{6}$, $\frac{R_\textrm{F}}{10}$},
        ticklabel style={font=\footnotesize},
        grid=major,
        scaled y ticks=false,              
        y tick label style={/pgf/number format/fixed}, 
        title={$N = 64$, $\eta = 0.0011$},
        title style={font=\footnotesize, yshift=-2mm},
    ]
        \addplot [
            blue, only marks, mark=*,
            mark size=1.7pt
        ] table[x index=0, y index=1, col sep=space] {figures/Error_var_N64.tex};

        \addplot [
            red, thick, mark=none
        ] table[x index=0, y index=2, col sep=space] {figures/Error_var_N64.tex};
    \end{axis}
    \end{tikzpicture}
    \hspace{-2mm}
    \begin{tikzpicture}
    \begin{axis}[
        width=4.25cm, height=5cm,
        ymin=0, ymax=1.6,
        xmin=2.41935, xmax=12.09675,
        xlabel={${r}_{1}$},
        label style={font=\footnotesize},
        xtick={12.09675, 8.0645, 6.048375, 4.0323, 2.41935},
        xticklabels={$\frac{R_\textrm{F}}{2}$, $\frac{R_\textrm{F}}{3}$, $\frac{R_\textrm{F}}{4}$, $\frac{R_\textrm{F}}{6}$, $\frac{R_\textrm{F}}{10}$},
        ticklabel style={font=\footnotesize},
        legend style={at={(-0.3,0.9)}, anchor=north west, font=\scriptsize, inner sep=1pt, fill opacity=0.75, draw opacity=1, text opacity=1},
        legend cell align=left,
        grid=major,
        title={$N = 128$, $\eta = 7.31\times10^{-5}$},
        title style={font=\footnotesize, yshift=-2mm},
    ]
        \addplot [
            blue, only marks, mark=*,
            mark size=1.7pt
        ] table[x index=0, y index=1, col sep=space] {figures/Error_var_N128.tex};
        \addlegendentry{Data}

        \addplot [
            red, thick, mark=none
        ] table[x index=0, y index=2, col sep=space] {figures/Error_var_N128.tex};
        \addlegendentry{Approx. in \eqref{eq:zeta}}
    \end{axis}
    \end{tikzpicture}

    \caption{Variance of the estimated distance versus true~distance, with $K \!=\! 1$, $f_\textrm{c} \!=\! 100$\!~GHz, $\textrm{SNR} \!=\! 20$\!~dB, $\theta_1 \! = \! 0$, and $d \!=\! \frac{\lambda_\textrm{c}}{2}$.} \vspace{-1mm}
    \label{Error_STD}
\end{figure}

\subsection{Impact of Interference on the Distance Estimation} \label{sec:MUSIC_2}

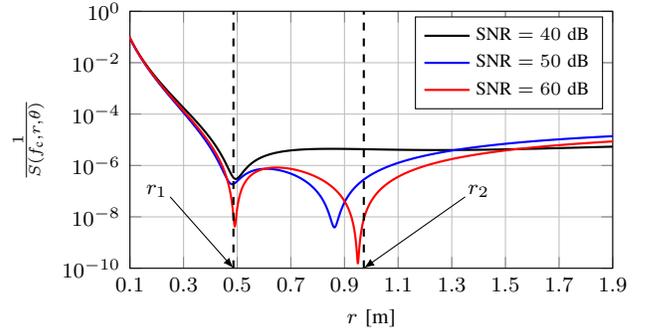
\begin{figure}[t!] 
    \centering
    \begin{tikzpicture}
        \begin{semilogyaxis}[
            width=8.5cm, height=5cm,
            xmin=0.1, xmax=1.9,
            ymin=1e-10, ymax=1,
            xlabel={$r$ [m]},
            ylabel={$\frac{1}{S(f_\textrm{c}, r, \theta)}$},
            x label style={font=\footnotesize},
            y label style={font=\footnotesize},
            ticklabel style={font=\footnotesize},
            grid=major,
            xtick={0.1,0.3,...,1.9},
            ytickten={-10,-8,...,0},
            ticklabel style={font=\footnotesize},
            legend style={at={(0.99,0.99)}, anchor=north east, font=\scriptsize},
        ]
            
            \addplot[black, thick] table[x index=0, y index=2, col sep=space, trim cells=true] {figures/equ28.tex};
            \addlegendentry{$\textrm{SNR} = 40$~dB}
            
            \addplot[blue, thick] table[x index=0, y index=3, col sep=space, trim cells=true] {figures/equ28.tex};
            \addlegendentry{$\textrm{SNR} = 50$~dB}
            
            \addplot[red, thick] table[x index=0, y index=4, col sep=space, trim cells=true] {figures/equ28.tex};
            \addlegendentry{$\textrm{SNR} = 60$~dB}

            \addplot[dashed, black, thick] coordinates {(0.486,1e-10) (0.486,1)};
            \addplot[dashed, black, thick] coordinates {(0.972,1e-10) (0.972,1)};

            \begin{scope}[>=latex]
                \draw[->] (axis cs:0.25,6e-8) -- (axis cs:0.486,1e-10);
            \end{scope}
            \node[black, font=\footnotesize] at (axis cs:0.2,10^-7) {$r_1$};
            
            \begin{scope}[>=latex]
                \draw[->] (axis cs:1.35,6e-8) -- (axis cs:0.972,1e-10);
            \end{scope}
            \node[black, font=\footnotesize] at (axis cs:1.4,1e-7) {$r_2$};

        \end{semilogyaxis}
    \end{tikzpicture}
    \caption{Denominator of the spectrum versus distance, with $K\!=\!2$, $N\!=\!3$, $f_{\textrm{c}} \!=\! 100$\!~GHz, $d \!=\! 9\lambda_\textrm{c}$, $r_{1} \!=\! \frac{R_\textrm{F}}{4}$, $r_{2} \!=\! \frac{R_\textrm{F}}{2}$, and $\theta_1 \!=\! \theta_2 \!=\!~0$.} \vspace{-1mm}
    \label{Spectrum_2UEs}
\end{figure}

To analyze the impact of interference on the distance estimation, we consider a simplified setup with $K = 2$ and $N = 3$, which represents the simplest scenario encompassing interference in the \ac{2D-MUSIC} algorithm. We assume that the two users are aligned in the angular domain, i.e., $\theta_{1} = \theta_{2}$. We begin by writing the denominator of \eqref{spectrum} as
\begin{align}\label{3Antenna2UEs}
    \hspace{-2mm} \frac{1}{S(f_\textrm{c}, r,\theta)}&=|u_1 b_1^* + u_2 b_2^* + u_3 b_3^*|^2\nonumber\\
    &=1+2|u_1||u_2|\cos\left(\frac{2\pi}{\lambda_\textrm{c}}(r-\bar{r}_0)+\phi_2 - \phi_1\right)\nonumber\\
    & \phantom{=} \ + 2|u_1||u_3|\cos\left(\frac{2\pi}{\lambda_\textrm{c}}(\bar{r}_2-\bar{r}_0)+\phi_3 - \phi_1\right)\nonumber\\
    & \phantom{=} \ + 2|u_2||u_3|\cos\left(\frac{2\pi}{\lambda_\textrm{c}}(\bar{r}_2-r)+\phi_3 - \phi_2\right),
\end{align}
with $\U_\textrm{n} = [u_1 \ u_2 \ u_3]^\tran \in \Compl^{3 \times 1}$, $\bar{r}_0 = \sqrt{r^2+d^2+2rd \sin \theta}$, and $\bar{r}_2 = \sqrt{r^2+d^2-2rd \sin \theta}$, and where $\phi_k = \angle u_k$ denotes the phase of $u_k$, $\forall k \in \{1, 2, 3\}$. While straightforward closed-form expressions for $\hat{r}_{1}$ and $\hat{r}_{2}$ cannot be derived, the minima of \eqref{3Antenna2UEs} accurately estimate the distances of the two users for sufficiently high \ac{SNR}. Fig.~\ref{Spectrum_2UEs} plots the denominator of the spectrum in \eqref{3Antenna2UEs} (evaluated numerically) versus the distance for different \acp{SNR} of user~$1$ (i.e., the closer user), with $f_{\textrm{c}} = 100$~GHz, $d = 9\lambda_\textrm{c}$, $r_{1} = \frac{R_\textrm{F}}{4}$, and $r_{2} = \frac{R_\textrm{F}}{2}$.\footnote{Here, we set $\sigma^2 = 1$ and normalize the channels with respect to the norm of the channel of user~$1$.} It is observed that a relatively high \ac{SNR} is required to accurately estimate the distance of user~$2$ (i.e., the farther user), since the spectrum is dominated by the signal of user~$1$ and the number of antennas is limited. For example, distance estimation for user~$2$ fails with $\textrm{SNR} = 40$~dB, whereas increasing the \ac{SNR} to $50$~dB and $60$~dB yields increasingly accurate estimates. Finally, increasing the number of antennas at the \ac{BS} alleviates the need for high \ac{SNR}, as it enhances spatial resolution and improves the separability of the users' signals.

\section{Sum-SE Evaluation} \label{sec:SE_evaluation}

In this section, we evaluate the \ac{UL} \ac{sum-SE} performance of the proposed localization-based beam focusing in the near field, examining the impact of key parameters such as the \ac{SNR} and carrier frequency. The results illustrate the advantages of employing the receive combiner in \eqref{eq:BF} over conventional \ac{ZF} with pilot-based channel estimation. While in Sections~\ref{sec:MUSIC_1} and \ref{sec:MUSIC_2} we focused on distance estimation, in the following we adopt the \ac{2D-MUSIC} algorithm to estimate both the distances and angles of the users.

We consider a \ac{BS} equipped with a \ac{ULA} comprising $N = 512$ antennas with spacing $d = \frac{\lambda_\textrm{c}}{2}$. $K = 2$ users transmitting with equal power, i.e., $\rho_1 = \rho_2$, are located at distances $r_1 = \frac{R_\textrm{F}}{8}$ and $r_2 = \frac{R_\textrm{F}}{2}$ (with $R_\textrm{F}\simeq 392$~m at $f_\textrm{c} = 100$~GHz) and angles $\theta_1 = \theta_2 = 0$ (i.e., aligned along the \ac{ULA}’s broadside direction). In addition to the \ac{LoS} path, we assume $L = 2$ clusters of scatterers uniformly distributed between the \ac{BS} and the users. Without loss of generality, the \ac{NLoS} paths are subject to the same reflection coefficient derived from linear fitting of the real measurement data in \cite{Con2025_Absorption} (Mortar~A, corresponding to a rough material), which yields $\alpha_{l}(f_\textrm{c}) = -0.0094\big(\frac{f_\textrm{c}}{10^9}\big) - 8.18$~dB. The length of the coherence block is set as $T = 5000$ at $f_\textrm{c} = 100$~GHz, which can accommodate a coherence bandwidth of $100$~MHz and a user speed of $15$~m/s. Moreover, $T$ scales proportionally with $\lambda_\textrm{c}$ as per \cite[Rem.~2.1]{Emil2017_mMIMO}. For pilot-based channel estimation, we adopt the \ac{LS} estimator with pilots drawn from the $\tau_{\textrm{Pil}}$-dimensional discrete Fourier transform matrix~\cite[Ch.~3.1.1]{Emil2017_mMIMO} with pilot length $\tau_{\textrm{Pil}} = 0.2T$, which offers a favorable trade-off between \ac{SINR} and pre-log factor of the \ac{sum-SE} across the whole range of considered parameters. For user localization via the \ac{2D-MUSIC} algorithm, we assume distance and angle steps $r_\textrm{step} \in \{10\lambda_\textrm{c}, 100\lambda_\textrm{c}\}$ and $\theta_\textrm{step} = 0.5^\circ$, respectively, with pilot length $\tau_{\textrm{Loc}} = 0.005T$. In this regard, setting $r_\textrm{step} = 10\lambda_\textrm{c}$ improves the localization accuracy at the cost of $10 \times$ higher computational complexity compared with $r_\textrm{step} = 100\lambda_\textrm{c}$. The plots are obtained by averaging over $100$ independent channel realizations.

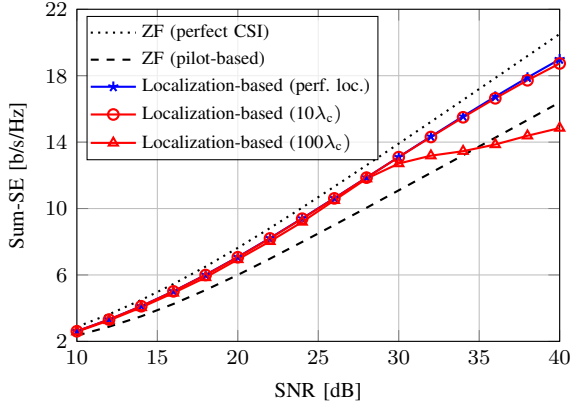
\begin{figure}[t]
    \centering
    \begin{tikzpicture}
    \begin{axis}[
        width=8.5cm, height=6cm,
        ymin=2, ymax=22,
        xmin=10, xmax=40,
        xtick={10,15,...,40},
        ytick={2,6,...,22},
        xlabel={SNR [dB]},
        ylabel={Sum-SE [b/s/Hz]},
        xlabel near ticks,
        ylabel near ticks,
        x label style={font=\footnotesize},
        y label style={font=\footnotesize},
        ticklabel style={font=\footnotesize},
        grid=major,
        legend style={at={(0.01,0.99)}, anchor=north west, font=\scriptsize, inner sep=1pt, fill opacity=0.75, draw opacity=1, text opacity=1},
        legend cell align=left,
    ]
    
    \addplot [black, dotted, thick]
    table[x index=0, y index=1, col sep=space] {figures/SE_vs_SNR_new.tex};
    \addlegendentry{ZF (perfect CSI)}

    \addplot [black, dashed, thick]
    table[x index=0, y index=4, col sep=space] {figures/SE_vs_SNR_new.tex};
    \addlegendentry{ZF (pilot-based)}
    
    \addplot [blue, thick, mark=star]
    table[x index=0, y index=2, col sep=space] {figures/SE_vs_SNR_new.tex};
    \addlegendentry{Localization-based (perf. loc.)}

    \addplot [red, thick, mark=o]
    table[x index=0, y index=5, col sep=space] {figures/SE_vs_SNR_new.tex};
    \addlegendentry{Localization-based ($r_\textrm{step} \! = \! 10 \lambda_\textrm{c})$}

    \addplot [red, thick, mark=triangle]
    table[x index=0, y index=3, col sep=space] {figures/SE_vs_SNR_new.tex};
    \addlegendentry{Localization-based ($r_\textrm{step} \! = \! 100 \lambda_\textrm{c})$}
    
    \end{axis}
    \end{tikzpicture}
    \caption{Sum-SE versus SNR of user~$1$ with $K = 2$, $N = 512$, $f_{\textrm{c}} = 100$~GHz, $d = \frac{\lambda_\textrm{c}}{2}$, $r_{1} = \frac{R_\textrm{F}}{8}$, and $r_{2} = \frac{R_\textrm{F}}{2}$.} \vspace{-1mm}
    \label{SE_vs_SNR}
\end{figure}

\begin{figure}[t]
    \centering
    \begin{tikzpicture}
    \begin{axis}[
        width=8.5cm, height=6cm,
        ymin=8, ymax=18,
        xmin=10, xmax=100,
        xtick={10,25,...,100},
        ytick={8,10,...,18},
        xlabel={$f_{\textrm{c}}$ [GHz]},
        ylabel={Sum-SE [b/s/Hz]},
        xlabel near ticks,
        ylabel near ticks,
        x label style={font=\footnotesize},
        y label style={font=\footnotesize},
        ticklabel style={font=\footnotesize},
        grid=major,
        legend style={at={(0.99,0.99)}, anchor=north east, font=\scriptsize, inner sep=1pt, fill opacity=0.75, draw opacity=1, text opacity=1},
        legend cell align=left,
    ]
    
    \addplot [black, dotted, thick]
    table[x index=0, y index=1, col sep=space] {figures/SE_vs_Frequency_new.tex};
    \addlegendentry{ZF (perfect CSI)}

    \addplot [black, dashed, thick]
    table[x index=0, y index=4, col sep=space] {figures/SE_vs_Frequency_new.tex};
    \addlegendentry{ZF (pilot-based)}
    
    \addplot [blue, thick, mark=star]
    table[x index=0, y index=2, col sep=space] {figures/SE_vs_Frequency_new.tex};
    \addlegendentry{Localization-based (perf. loc.)}

    \addplot [red, thick, mark=o]
    table[x index=0, y index=5, col sep=space] {figures/SE_vs_Frequency_new.tex};
    \addlegendentry{Localization-based ($r_\textrm{step} \! = \! 10 \lambda_\textrm{c})$}

    \addplot [red, thick, mark=triangle]
    table[x index=0, y index=3, col sep=space] {figures/SE_vs_Frequency_new.tex};
    \addlegendentry{Localization-based ($r_\textrm{step} \! = \! 100 \lambda_\textrm{c})$}
    
    \end{axis}
    \end{tikzpicture}
    \caption{Sum-SE versus carrier frequency with $K = 2$, $N = 512$, $\rho_1 = \rho_2 = 23$~dBm, $d = \frac{\lambda_\textrm{c}}{2}$, $r_{1} = \frac{R_\textrm{F}}{8}$, and $r_{2} = \frac{R_\textrm{F}}{2}$.} \vspace{-1mm}
    \label{SE_vs_frequency}
\end{figure}
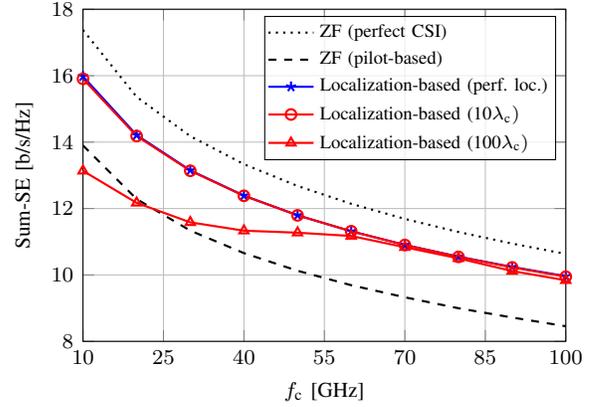

Considering $f_\textrm{c} = 100$~GHz, Fig.~\ref{SE_vs_SNR} plots the \ac{sum-SE} as a function of the \ac{SNR} of user~$1$ (i.e., the closer user), obtained by setting $\sigma^2 = 1$ and normalizing the channels with respect to the norm of the channel of user~$1$. The proposed localization-based beam focusing with distance step $r_\textrm{step} = 100\lambda_\textrm{c}$ outperforms pilot-based \ac{ZF} except at very high \ac{SNR} (over $35$~dB): in this regime, the coarse distance resolution becomes the dominant source of error, and further increasing the \ac{SNR} no longer improves the localization accuracy. Increasing the distance resolution by setting $r_\textrm{step} = 10\lambda_\textrm{c}$ yields a \ac{sum-SE} close to that with perfect localization, which in turn approaches the performance of \ac{ZF} with perfect \ac{CSI}. Fig.~\ref{SE_vs_frequency} illustrates the impact of the carrier frequency on the \ac{sum-SE}. Note that, since $r_\textrm{step}$ scales with the wavelength, a distance step of $100\lambda_\textrm{c}$ yields a significant localization error at low frequencies, thereby penalizing localization-based beam focusing. Here, the \ac{AWGN} power is computed as $\sigma^2 = \xi - 174 + 10\log_{10}(B)$~[dBm], where $\xi = 13$~dB is the noise figure and $B$~[Hz] denotes the communication bandwidth, which is set as $B = 0.001 f_{\textrm{c}}$ (e.g., $B = 100$~MHz at $f_{\textrm{c}} = 100$~GHz) and matches the coherence bandwidth. As the frequency grows and the channels becomes increasingly \ac{LoS}-dominated (see Section~\ref{Near-Field Channel Model}), the proposed localization-based beam focusing remains effective under shorter coherence blocks and higher \ac{AWGN} power. Specifically, as $T$ (and thus $\tau_{\textrm{Pil}}$) decreases and $\sigma^{2}$ increases, the accuracy of pilot-based channel estimation deteriorates, whereas localization-based beam focusing is less affected by the reduced $\tau_{\textrm{Loc}}$ and stronger \ac{AWGN}. Hence, despite the added offline complexity of \ac{2D-MUSIC}, beam-focusing design based on user localization represents a viable alternative to pilot-based methods in \ac{LoS}-dominated \ac{mmWave} and \ac{sub-THz} systems.

\vspace{-1mm}

\section{Conclusions}

\vspace{-1mm}

We proposed a localization-based beam-focusing design \linebreak

\noindent for near-field communications and analyzed the \ac{2D-MUSIC} algorithm in terms of distance estimation error and interference effects. We evaluated the proposed design in terms of \ac{UL} \ac{sum-SE}, showing that it is generally superior to pilot-based \ac{ZF} under \ac{LoS}-dominated propagation typical of high-frequency systems. Future work will focus on analytically characterizing the \ac{sum-SE} performance and optimizing the receive combiners under imperfect location estimates.

\bibliographystyle{IEEEtran}
\bibliography{refs_abbr,refs}

\end{document}